\documentclass{eas}
\pdfoutput=1
\usepackage{graphicx}
\usepackage{multirow}
\usepackage[varg]{txfonts} 
\usepackage[latin1]{inputenc}
\usepackage{textcomp}
\usepackage{amssymb}
\newcommand{\Msolar}{M$_{\odot}$ }
\newcommand{\Simname}{\includegraphics[height=0.3cm]{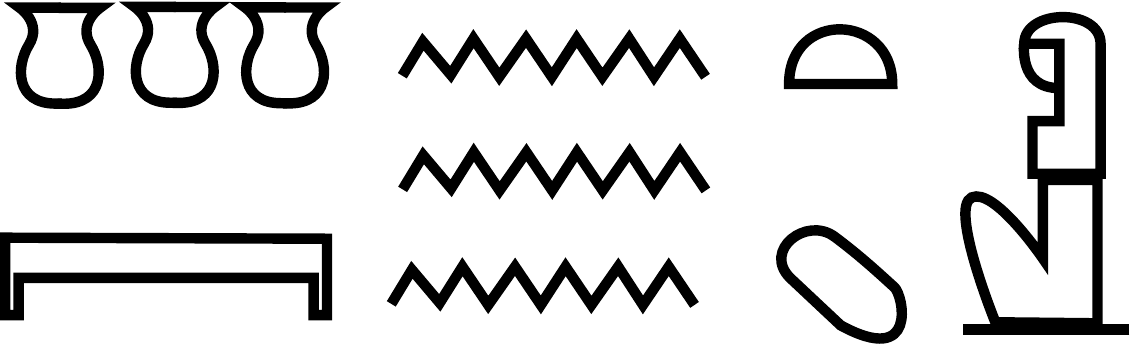} (Nut) }

\newcommand{\Nutref}{({Powell:2010inprep}, in prep) }
\TitreGlobal{A Universe of Dwarf Galaxies, Lyon 2010}
\begin{document}

\title{How Does Feedback Affect Milky Way Satellite Formation?}
\author{Sam Geen}\address{Department of Physics, University of Oxford, Denys Wilkinson Building, Keble Road, Oxford, OX1 3RH, United Kingdom}
\author{Adrianne Slyz}
\author{Julien Devriendt}
\begin{abstract}We use sub-parsec resolution hydrodynamic resimulations of a Milky Way (MW) like galaxy at high redshift to investigate the formation of the MW satellite galaxies. More specifically, we assess the impact of supernova feedback on the dwarf progenitors of these satellite, and the efficiency of a simple instantaneous reionisation scenario in suppressing star formation at the low-mass end of this dwarf distribution. Identifying galaxies in our high redshift simulation and tracking them to z=0 using a dark matter halo merger tree, we compare our results to present-day observations and determine the epoch at which we deem satellite galaxy formation must be completed. We find that only the low-mass end of the population of luminous subhalos of the Milky-Way like galaxy is not complete before redshift 8, and that although supernovae feedback reduces the stellar mass of the low-mass subhalos (M$\le 10^{9}$\Msolar), the number of surviving satellites around the Milky-Way like galaxy at z = 0 is the same in the run with or without supernova feedback. If a luminous halo is able to avoid accretion by the Milky-Way progenitor before redshift 3, then it is likely to survive as a MW satellite to redshift 0.
\end{abstract}
\maketitle
\section{Introduction}
\label{intro}
The ``Missing Satellite Problem'' arose from the discovery that pure N-body simulations of MW-sized halos contain at least an order of magnitude more dark substructures than observed luminous satellites (\cite{Moore:1999p565,Klypin:1999p1871}). Although this discrepancy has recently been alleviated by the detection of a host of of new MW satellites (\cite{Koposov:2008p793,Tollerud:2008p1758}), a non negliglible (factor of a few) overabundance of dark substructures remains. Processes such as reionisation (\cite{Bullock:2000p1416,Gnedin:2000p1832,Benson:2002p1802,Somerville:2002p1760}) and supernova feedback (\cite{Dekel:1986p1417,Benson:2003p1159}) have been invoked to reduce the number of satellite galaxies by heating and expelling potentially star-forming gas from low-mass halos.

To test this claim, we are running extremely high resolution hydrodynamic cosmological resimulations of a MW-like galaxy and its surroundings, improving on the work of other authors (\cite{Okamoto:2009p1754,Wadepuhl:2010p1805}) by at least an order of magnitude in physical resolution.

In this paper, we address the issue of whether MW satellite formation was complete by the epoch of reionisation or not. In addition, since we run identical simulations, but with and without supernova feedback, we assess the role of supernovae during the formation of satellite galaxies at high redshift. Since our resimulations containing star formation do not reach z=0, we compare our results to observations of MW satellites by twinning the halos containing galaxies in these high-redshift simulations with halos in an adiabatic gas simulation run to z=0. Along the lines of \cite{Moore:2006p1747,Strigari:2007p801}, we track dark matter halos to z=0 to determine the properties of surviving halos. We then proceed to quantify the effect of various physical processes on dwarf galaxy formation at high redshift, and determine the importance of such processes in shaping the main properties of the MW satellite galaxies we see today.

\section{Numerical Simulations}
\label{simulations}

\begin{figure*}
\includegraphics[width=0.43\hsize]{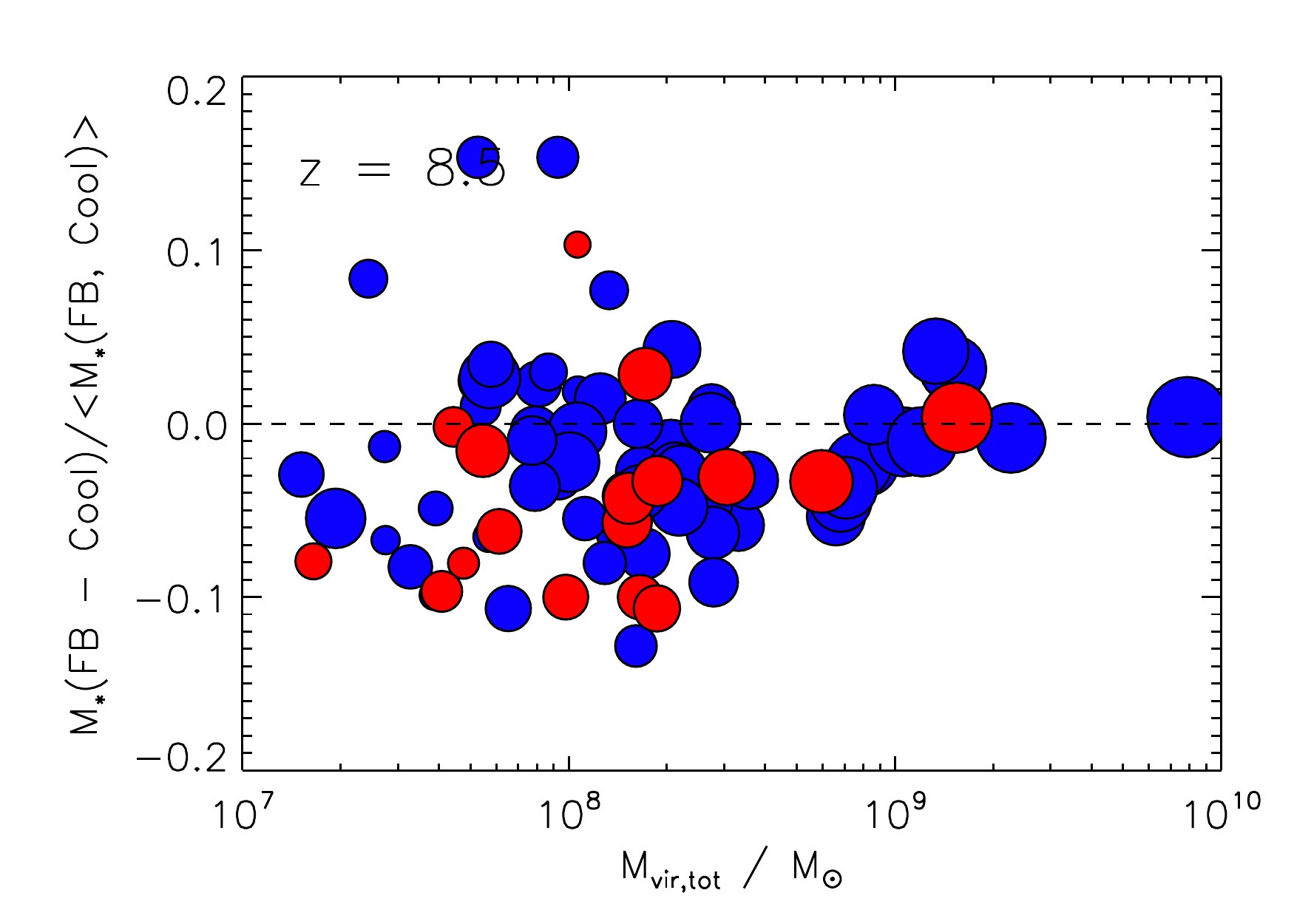}
\includegraphics[width=0.43\hsize]{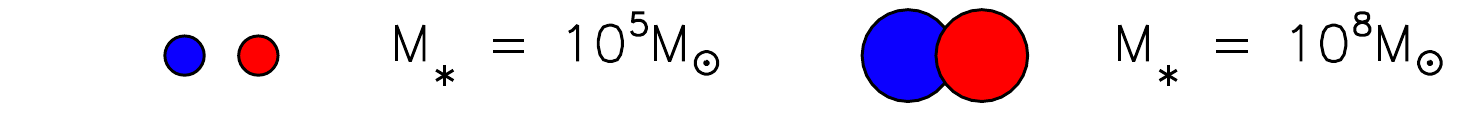}
 \caption{Comparison of stellar masses of galaxies at z=8.5 in the Cooling and Feedback run whose particles end up inside the MW virial radius at z=0, showing the fractional difference between the values in the two runs vs total virial mass in Adiabatic Run. A positive value indicates that the inclusion of supernova feedback enhances star formation, while a negative value means that it suppresses star formation. The circle size indicates stellar mass (see legend). Red circles indicate halos that survive as MW satellites to z=0; blue circles are completely disrupted by merger events by z=0.}
\label{starmass_compare}
\end{figure*}

In this work, we analyse three resimulations of a Milky Way-like galaxy from the \Simname suite of simulations \Nutref named ``Adiabatic'', ``Cooling'' and ``Feedback''. We carry out these simulations using the Adaptive Mesh Refinement code RAMSES (\cite{Teyssier:2002p533}). All runs contain dark matter and baryonic gas. For the Cooling and Feedback runs, we include metal dependent radiative energy losses by the gas and its heating by a spatially-uniform UV background (\cite{Haardt:1996p1457}) after z=8.5, along with star formation. The threshold density for star formation is $10^5$atoms/cm$^3$ with an efficiency of 1\%. For the Feedback run, we include supernova feedback, which also enriches the gas with metals. These simulations have a maximum physical resolution of 0.5pc, with a minimum mass of 5.6$\times10^{4}$\Msolar for dark matter particles and 76\Msolar for star particles (\cite{Geen:2011inprep}). With this mass resolution we are able to capture individual supernovae, although we still treat star particles as populations of stars rather than individual stars.

\section{Tracking Halos}
\label{halos}

We identify dark matter (sub)halos and clusters of stars using HaloMaker with the Most Massive Subhalo Method (\cite{Tweed:2009p996}). We then associate each cluster of stars to its nearest dark matter (sub)halo. For the adiabatic run, we use TreeMaker (\cite{Tweed:2009p996}) to build a halo merger tree of the MW-like halo from our initial conditions to z=0. We then attach all dark matter (sub)halos in the Cooling and Feedback runs to their nearest ``twins'' in the Adiabatic run (based on the number of dark matter particles shared between the two halos) for a selection of redshifts before and after reionisation between z=11 and z=8 (\cite{Geen:2011inprep}).
 
Using the Adiabatic run merger tree, we track each ``twinned'' halo to z=0 and find whether it has survived as an independent halo, as a subhalo of the MW, or has been completely disrupted in a merger with the MW or another halo. We posit that since all our galaxies are embedded within a dark matter (sub)halo, if this host halo survives to z=0 then so does the embedded galaxy.

From the 87 galaxies formed in the Feedback run at z=8.5 (just prior to reionisation) we find 13 galaxies survive as MW satellites to z=0, with 4 being destroyed in mergers with other satellite progenitor galaxies (6 in the Cooling run) and 71 destroyed in mergers with the MW (69 in the Cooling run). At z=8 (51 Myr after reionisation), one further satellite galaxy progenitor is formed that survives to z=0.

\section{Feedback and Reionisation}
\label{feedback_rei}

\begin{figure*}
\centerline{\includegraphics[width=0.45\hsize]{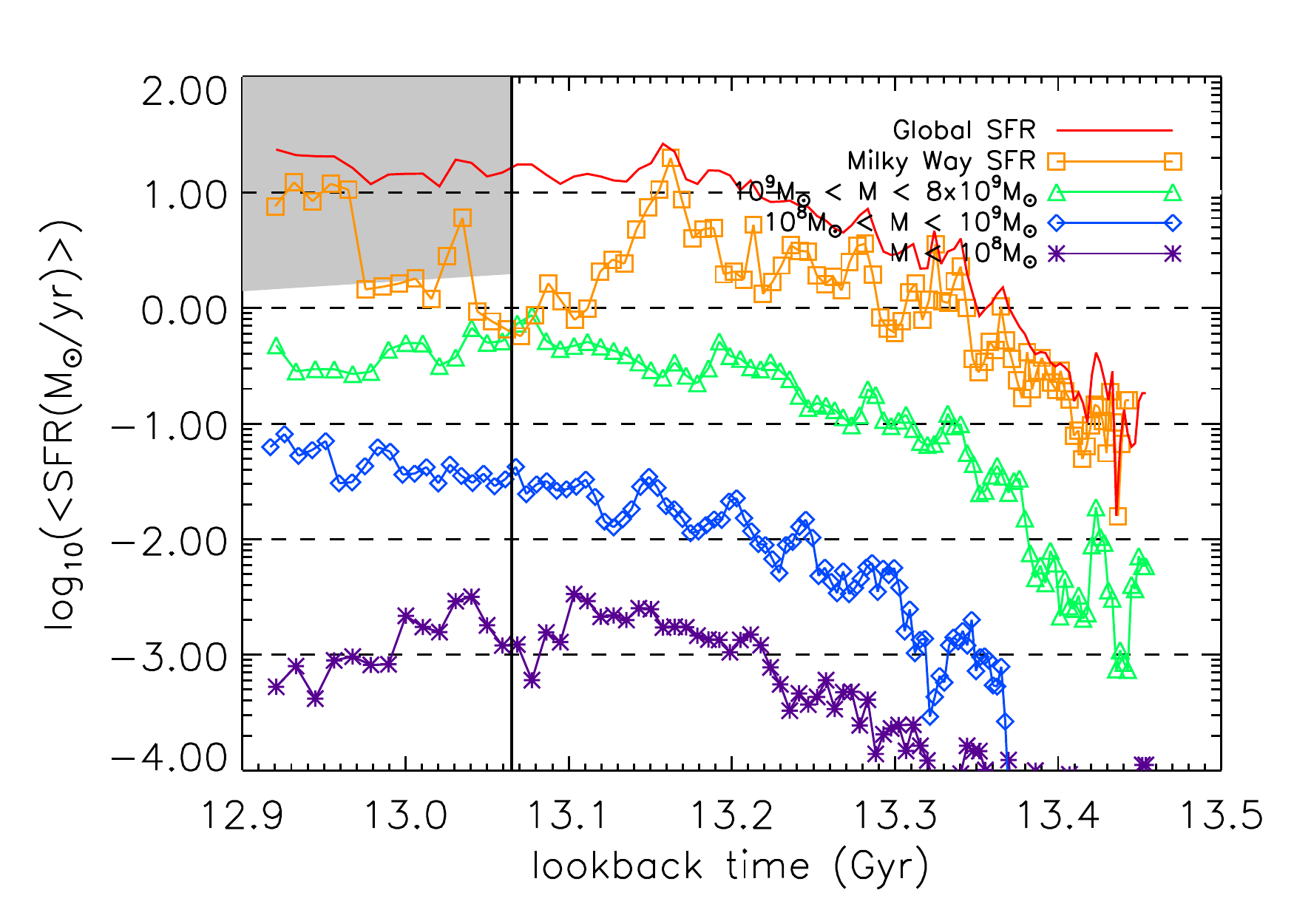} \includegraphics[width=0.45\hsize]{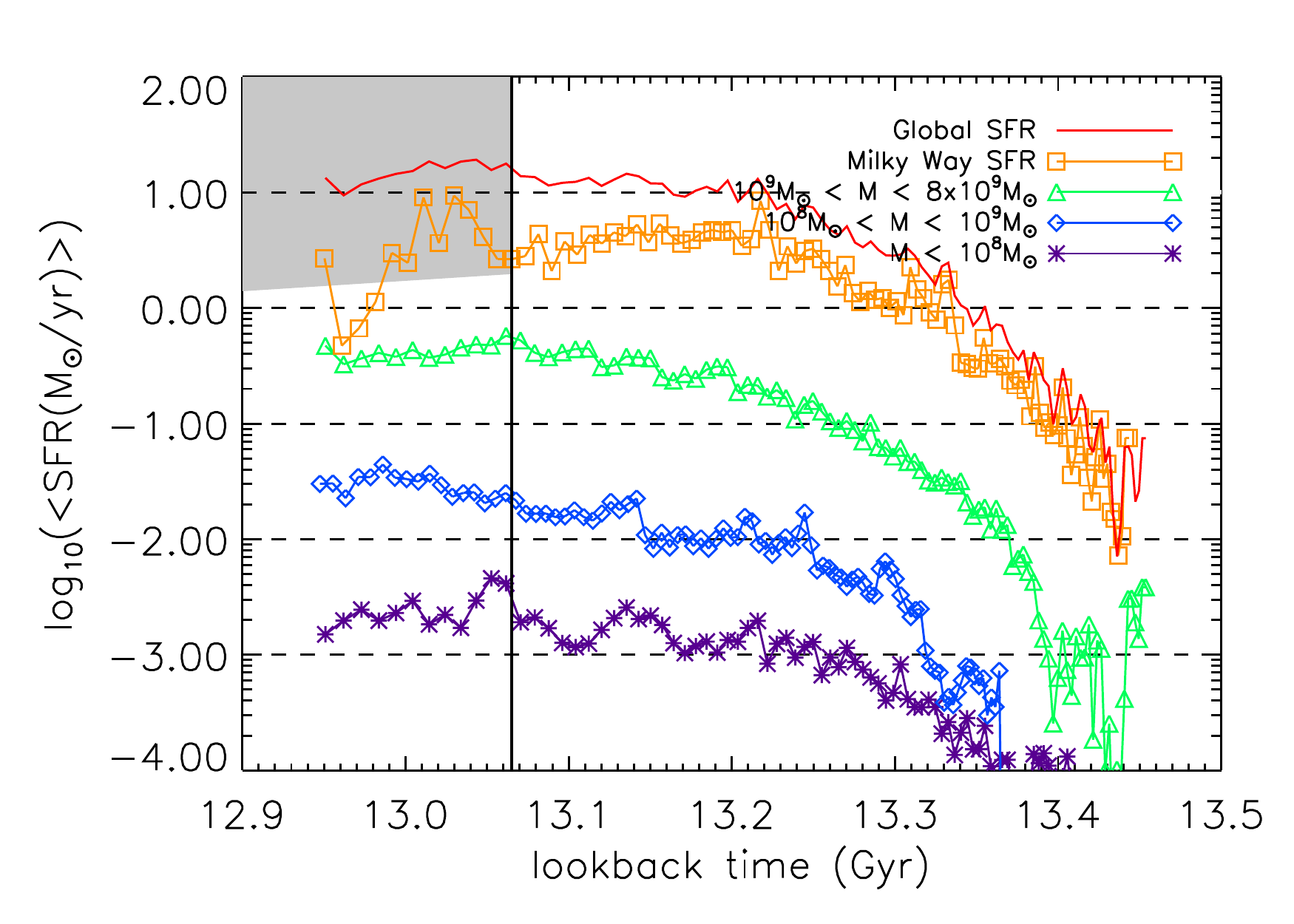}}
  \caption{Star formation histories in the Cooling run (left) and Feedback run (right). The vertical line at z=8.5 (look-back time 13.065Gyr) shows the point at which reionisation is turned on in the simulations. The grey region shows the detectable star formation rates as determined by \protect\cite{Wilkins:2010p1806}.}
\label{sfrplots}
\end{figure*}

Comparing the Cooling and Feedback runs, we find that galaxies with total mass under $10^9$\Msolar generally exhibit suppressed star formation when our model of supernova feedback is included. By contrast, galaxies with total mass over $10^9$\Msolar have similar masses in both runs, with some galaxies showing signs of enhanced star formation when supernova feedback is included (figure \ref{starmass_compare}).

The latter is a consequence of enhanced cooling and hence higher star formation rate due to the introduction of metals from supernovae into the interstellar medium. However, supernova feedback does not appear to affect the number of galaxies formed, with roughly similar numbers of galaxies formed in both the Cooling and Feedback run. Further, the same MW satellites are found in both the Cooling and Feedback run when tracked to z=0 as described in section \ref{halos}.

We find that reionisation does not immediately halt star formation, even in the smallest halos, in agreement with \cite{Wadepuhl:2010p1805} (see figure \ref{sfrplots}). There appears to be a slight downward trend in the star formation rate after reionisation, however. As the simulations continue beyond 200Myr after reionisation, we will monitor whether this trend continues. Since we implement a uniform ionising background, our galaxies cannot self-shield. We plan to address this issue in future studies by including radiative transfer of hydrogen-ionising radiation.


\section{MW Satellites Today}
\label{sats_today}

\begin{figure*}
\centerline{\includegraphics[width=0.43\hsize]{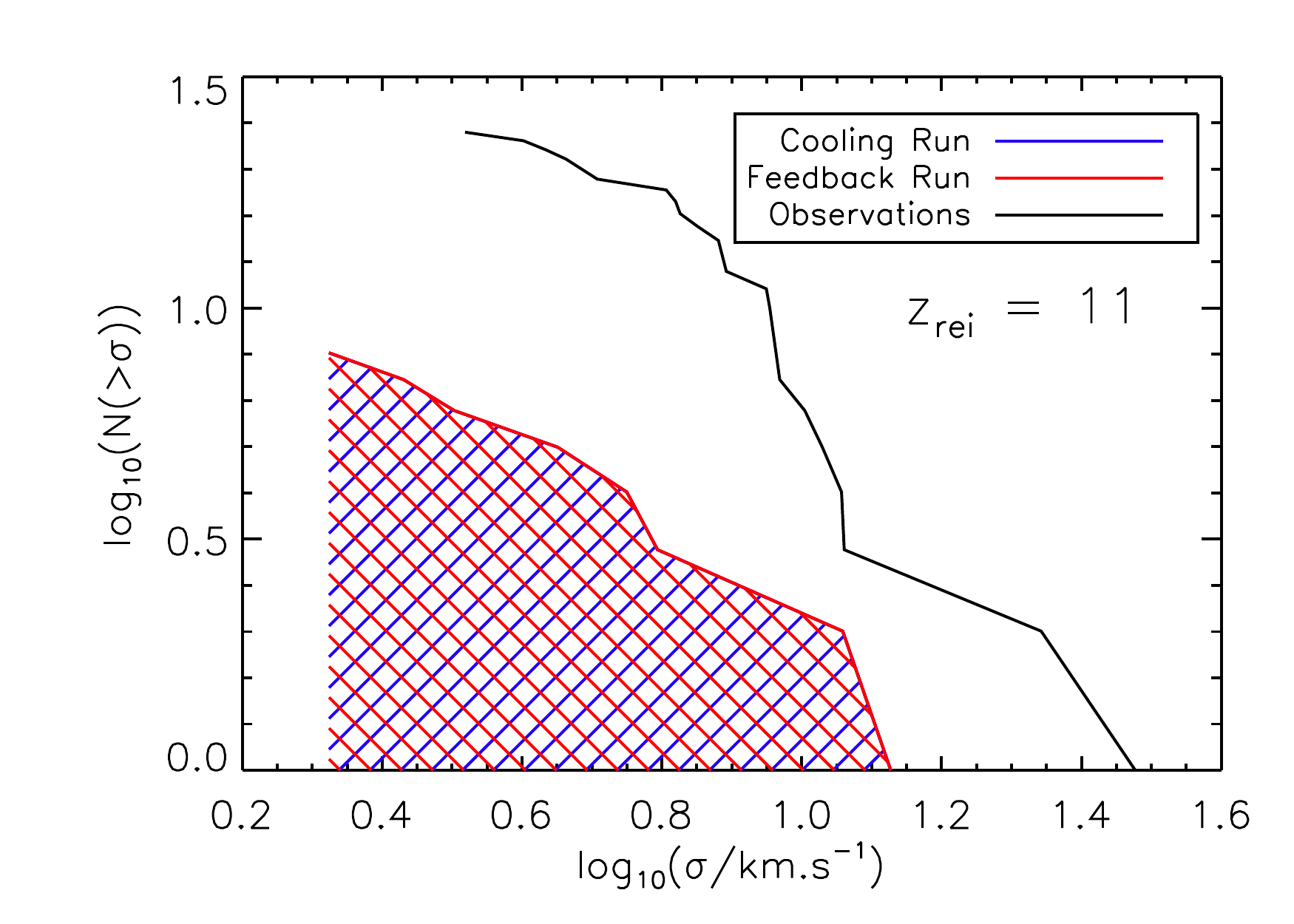}
\includegraphics[width=0.43\hsize]{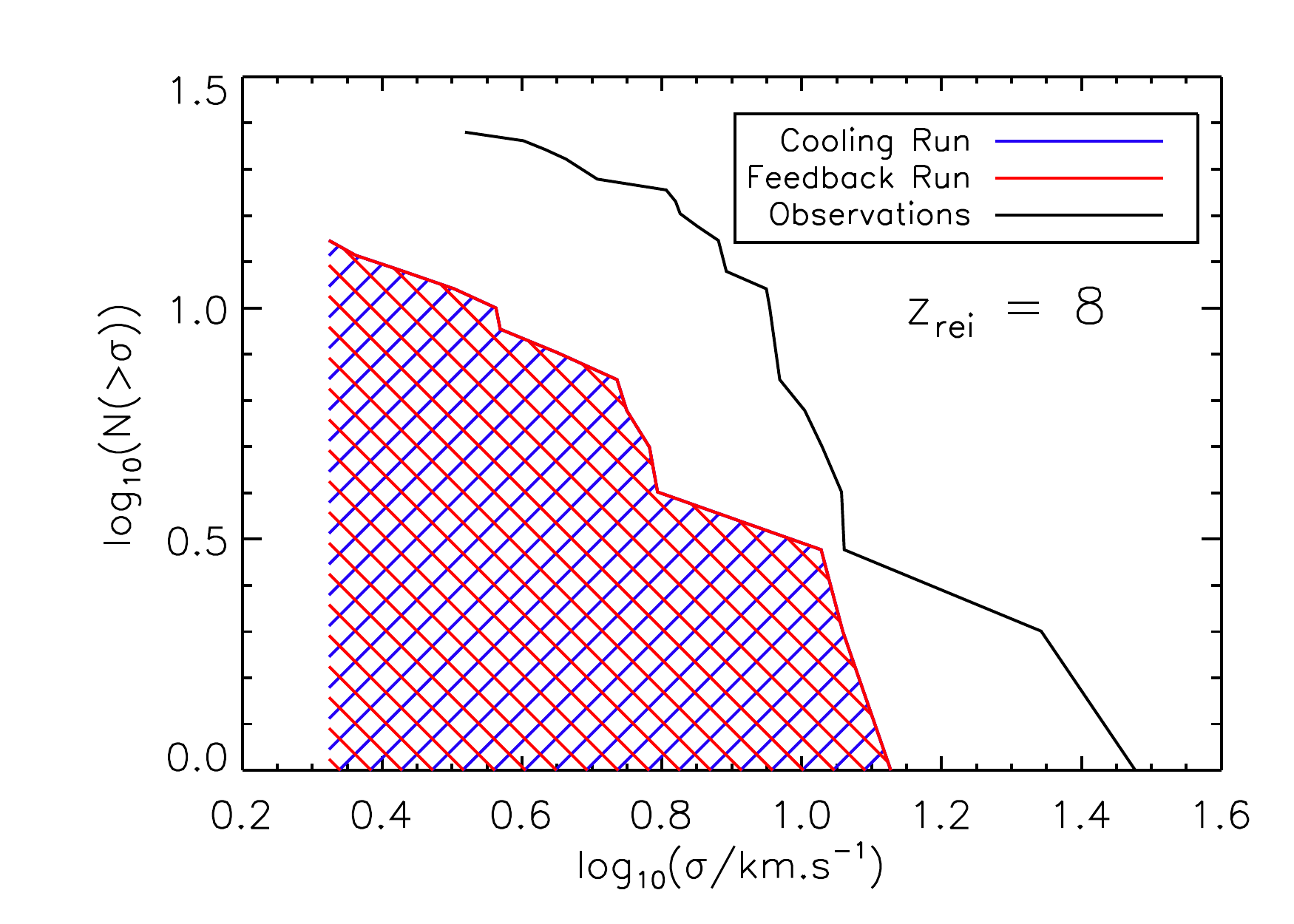}}
 \caption{Cumulative velocity dispersion distributions of halos with galaxies and/or star clusters embedded in the Cooling (Blue) and Feedback (Red) runs, tracked to z=0 as explained in section \protect\ref{halos}. Observations (black line) use data from (\protect\cite{Mateo:1998p773,Bekki:2005p1869,Bekki:2009p1567,Wolf:2010p1868}). The simulated data is given as the maximum circular velocity / $\sqrt2$. The error bars represent the sampling error on the data in each bin.}
\label{veldisp_stars}
\end{figure*}

As described in section \ref{halos}, we track the halos containing galaxies in the Cooling and Feedback runs to z=0 using the Adiabatic run merger tree. We find that the most important criterion for a halo to survive to z=0 is its redshift of capture, with halos captured before z=3 being completely disrupted by z=0. However, this redshift value may be dependent on the host mass.



In figure \ref{veldisp_stars}, we compare the observed stellar velocity dispersions to our simulated galaxies' dark matter velocity dispersions at z=0 in the Adiabatic run, since estimates of virial mass from observations have typically large errors \cite{Wolf:2010p1868}. For all sampled redshifts (z=11 to z=8), we find too few satellites when compared with the observed MW satellites, with better agreement if we wait until lower redshift before sampling the galaxies. Assuming this trend continues, this suggests that satellite galaxy formation is not complete by reionisation. Indeed, we find one galaxy formed between z=8.5 and z=8, after our implemented epoch of reionisation. Furthermore, \cite{Koposov:2008p793,Tollerud:2008p1758} suggest that the current sample of observed MW satellites is incomplete. Since our runs containing star formation have not reached z=0, we cannot comment on the completeness of our simulations.

\section{Conclusions}
\label{conclusions}

We find that the velocity dispersions of satellites we produce underproduces those of observed Milky Way satellites, suggesting that satellite progenitor formation is not complete by reionisation. Star and galaxy formation continue past reionisation, with little sign of reionisation immediately halting the star formation rate. Supernova feedback reduces the star formation in halos with total mass $ < 10^9$\Msolar, but not the number of galaxies formed. Finally, we find only halos captured after z=3 survive to z=0 as Milky Way satellites.

\textbf{Acknowledgements} The authors would like to thank Romain Teyssier for useful comments and discussions during the work presented here. The simulations were performed on the JADE supercomputer at CINES, France.



\end{document}